\documentclass{article}
\usepackage{graphicx}
\input{tcilatex}

\begin{document}

\begin{center}
{\Large Functional Integral Method in the Theory }

%TCIMACRO{
%\TeXButton{TeX field}{\vskip 0.2cm
%}}
%BeginExpansion
\vskip 0.2cm
%
%EndExpansion
{\Large of Color Superconductivity}{\footnote{
Invited talk at the International Conference on Physics at Extreme Energy-Rencontres du
Vietnam IV, Hanoi 19-25 July 2000}}

%TCIMACRO{
%\TeXButton{TeX field}{\vskip 0.4cm
%}}
%BeginExpansion
\vskip 0.4cm
%
%EndExpansion
\textbf{Nguyen Van Hieu}

\textit{Institute of Physics, NCST of Vietnam }

and

\textit{Faculty of Technology, Vietnam National University, Hanoi}

%TCIMACRO{
%\TeXButton{TeX field}{\vskip 1cm
%}}
%BeginExpansion
\vskip 1cm
%
%EndExpansion
\textbf{Abstract}
\end{center}

\begin{quotation}
We propose to study the superconducting pairing of quarks with the formation
of the diquarks as well as the quark-antiquark pairing in QCD by means of
the functional integral technique. The dynamical equations for the
superconducting order parameters are the nonlinear integral equations for
the composite quantum fields describing the quark-quark or quark-antiquark
systems. These composite fields are the bi-local fields if the pairing is
generated by the gluon exchange while for the instanton induced pairing
interactions they are the local ones. The expressions of the free energy
densities are derived. The binding of three quarks is also discussed.
\end{quotation}

%TCIMACRO{\TeXButton{TeX field}{\vskip 0.5cm}}
%BeginExpansion
\vskip 0.5cm%
%EndExpansion
The superconducting pairing of quarks due to the gluon exchange in QCD with
the formation of the diquark condensate was proposed by Barrois$^{\left[
1\right] }$ and Frautshi$^{\left[ 2\right] }$ since more than two decades
and then studied by Bailin and Love$^{\left[ 3\right] }$, Donoglue and
Sateesh$^{\left[ 4\right] }$, Iwasaki and Iwado$^{\left[ 5\right] }.$
Recently, in a series of papers by Alford, Rajagopal and Wilczek$^{\left[
6\right] }$, Sch\"{a}fer and Wilczek$^{\left[ 7\right] }$, Rapp,
Sch\"{a}fer, Shuryak and Velkovsky$^{\left[ 8\right] }$, Evans, Hsu and
Schwetz$^{\left[ 9\right] }$, Son$^{\left[ 10\right] }$, Carter and Diakonov$%
^{\left[ 11\right] }$ and others$^{\left[ 12-21\right] }$ there arose a new
interest to the existence of the diquark Bose condensate in the QCD dense
matter-the color superconductivity. The connection between the color
superconductivity and the chiral phase transition in QCD was studied by
Berges and Rajagopal$^{\left[ 22\right] }$, Harada and Shibata$^{\left[
23\right] }$. There exists also the spontaneous parity violation, as it was
shown by Pisarski and Rischke$^{\left[ 24\right] }$.

For the study of many-body systems of relativistic particles with the
internal degrees of freedom as well as with the virtual creation and
annihilation of the particle-antiparticle pairs the functional integral
technique is a powerful mathematical tool. This method was applied to the
study of the color superconductivity as well as the quark-antiquark pairing
in a series of papers of the group of authors in Hanoi$^{\left[ 25-28\right]
}$. In my talk I present a review of these works.

Denote $\psi _{A}$ the quark field, where $A=\left( \alpha ,a,i\right) $ is
the set consisting of the Dirac spinor index $\alpha =1,2,3,4$ and the color
and flavor indices $a=1,2,3...N_{c}$ and $i=1,2,3,....N_{f}$. The internal
symmetry groups are assumed to be $SU\left( N_{c}\right) _{c}$ and $SU\left(
N_{f}\right) _{f}$.We work in the imaginary time formalism and denote $%
x=\left( \tau ,\mathbf{x}\right) $ a vector in the Euclidean
four-dimensional space-time. The partition function of the system of free
quarks and antiquarks with the chemical potential $\mu $ and at the
temperature $T$ can be expressed in the form of the functional integral

\begin{equation}
Z_{0}=\int \left[ D\psi \right] \left[ D\overline{\psi }\right] \exp \left\{
-\int dx\overline{\psi }^{A}\left( x\right) D_{A}^{B}\psi _{B}\left(
x\right) \right\}  \label{1}
\end{equation}
with brief notations

\begin{equation}
\int dx=\int\limits_{0}^{\beta }d\tau \int d\mathbf{x,\qquad }\beta =\frac{1%
}{kT},\qquad D_{A}^{B}=\left[ \gamma _{4}\left( \frac{\partial }{\partial
\tau }-\mu \right) +\mathbf{\gamma \nabla }+M\right] _{A}^{B},  \label{2}
\end{equation}
where $k$ is the Boltzmann constant and $M$ is the bare quark mass. We start
from the study of the system with the direct instanton induced four-fermion
couplings and then consider the case of the pairing due to the gluon
exchange.

%TCIMACRO{\TeXButton{TeX field}{\vskip 0.5cm}}
%BeginExpansion
\vskip 0.5cm%
%EndExpansion
We write the expression of the interaction Lagrangian of the direct
four-fermion couplings in two different forms

\begin{equation}
L_{int}=\frac{1}{2}\overline{\psi }^{A}(x)\overline{\psi }%
^{C}(x)V_{CA}^{BD}\psi _{D}(x)\psi _{B}(x)  \label{3}
\end{equation}
and

\smallskip 
\begin{equation}
L_{int}=\frac{1}{2}\overline{\psi }^{A}(x)\psi _{B}(x)U_{AC}^{BD}\overline{%
\psi }^{C}(x)\psi _{D}(x)  \label{4}
\end{equation}
convenient for the study of the quark-quark or quark-antiquark pairing,
resp. The notations $V_{CA}^{BD}\hspace{0in}\;$and $U_{AC}^{BD}\;$are
related each to other

$\qquad $%
\[
V_{CA}^{BD}\hspace{0in}=U_{AC}^{BD}. 
\]
These coupling constants are antisymmetric under the interchanges of the
indices.

In order to study the quark-quark pairing we use the interaction Lagrangian
in the form (3) and have the partition function

\begin{eqnarray}
Z &=&\int [D\psi ][D\overline{\psi }]\exp \left\{ -\int dx\overline{\psi }%
^{A}(x)D_{A}^{B}\psi _{B}(x)\right\}  \nonumber \\
&&.\exp \left\{ \frac{1}{2}\int dx\overline{\psi }^{A}(x)\overline{\psi }%
^{C}(x)V_{CA}^{BD}\hspace{0in}\psi _{D}(x)\psi _{B}(x)\right\} .  \label{5}
\end{eqnarray}
Introduce the composite local bi-spinor fields $\Phi _{DB}(x),\overline{\Phi 
}^{AC}(x)$ describing the diquarks and consider the functional integral

\begin{equation}
Z_{0}^{\Phi ,\overline{\Phi }}=\int [D\Phi ][D\overline{\Phi }]\exp \left\{ -%
\frac{1}{2}\int dx\overline{\Phi }^{AC}(x)V_{CA}^{BD}\hspace{0in}\Phi
_{DB}(x)\right\} .  \label{6}
\end{equation}
Applying the corresponding Hubbard-Stratonovich transformation, we rewrite $%
Z $ in the form of the functional integral over these composite local fields

\begin{equation}
Z=\frac{Z_{0}}{Z_{0}^{\Phi ,\,\overline{\Phi }}}\int [D\Phi ][D\overline{%
\Phi }]\exp \left\{ S_{\text{eff}}[\Phi ,\overline{\Phi }]\right\} .
\label{7}
\end{equation}
The field equations

\begin{equation}
\frac{\delta S_{\text{eff}}[\Phi ,\overline{\Phi }]}{\delta \overline{\Phi }%
^{AC}(x)}=0  \label{8}
\end{equation}
are the dynamical equations for the quark-quark pairing. They can be written
in the following manner

\begin{eqnarray}
\Delta _{D_{1}B_{1}}(x_{1}) &=&V_{D_{1}B_{1}}^{A_{1}C_{1}}\hspace{0in}%
\left\{ \int dx_{2}S_{C_{1}}^{C_{2}}(x_{1}-x_{2})\Delta
_{C_{2}A_{2}}(x_{2})S_{A_{1}}^{A_{2}}(x_{1}-x_{2})\right.  \nonumber \\
&&\left. -\int dx_{2}\int dx_{3}\int
dx_{4}S_{C_{1}}^{C_{2}}(x_{1}-x_{2})\Delta
_{C_{2}A_{2}}(x_{2})S_{A_{3}}^{A_{2}}(x_{3}-x_{2})\right.  \nonumber \\
&&\left. .\,\overline{\Delta \,}^{A_{3}C_{3}}\left( x_{3}\right)
S_{C_{3}}^{C_{4}}\left( x_{3}-x_{4}\right) \Delta _{C_{4}A_{4}}\left(
x_{4}\right) S_{A_{1}}^{A_{4}}\left( x_{1}-x_{4}\right) +....,\right\}
\label{9}
\end{eqnarray}
where $S_{A}^{B}(x-y)$ are the propagators of the free quarks and

\begin{equation}
\Delta _{CA}(x)=V_{CA}^{BD}\Phi _{DB}(x),\qquad \overline{\Delta }%
^{BD}(x)=V_{CA}^{BD}\overline{\Phi }^{AC}(x).  \label{10}
\end{equation}
Consider the special class of the constant solutions

\[
\Delta _{CA}=const,\qquad \overline{\Delta }^{BD}=const. 
\]
Denote $\widetilde{S}_{A}^{B}(\mathbf{p},\varepsilon _{m})\;$the Fourier
transforms of the free quark propagators, $\varepsilon _{m}=\frac{\pi }{%
\beta }\left( 2m+1\right) $, and introduce $\widehat{\Delta },\,\widehat{%
\overline{\Delta }}\;$and $\widehat{S}(\mathbf{p},\varepsilon _{m})\;$the
matrices with the elements $\Delta _{CA},\overline{\Delta }^{BD}$ and $%
\widetilde{S}_{A}^{B}(\mathbf{p},\varepsilon _{m})$. Then we have following
matrix equation for the quark-quark pairing

\begin{eqnarray}
\left[ \widehat{\Delta }\right] _{DB} &=&V_{DB}^{AC}\hspace{0in}\frac{1}{%
\beta }\stackunder{m}{\sum }\frac{1}{(2\pi )^{3}}\int d\mathbf{p}  \nonumber
\\
&&\left[ \frac{1}{1+\widehat{S}(\mathbf{p},\varepsilon _{m})\widehat{\Delta }%
\widehat{S}^{T}(-\mathbf{p},-\varepsilon _{m})\widehat{\overline{\Delta }}}S(%
\mathbf{p},\varepsilon _{m})\widehat{\Delta }\widehat{S}^{T}(-\mathbf{p}%
,-\varepsilon _{m})\right] _{CA}  \label{11}
\end{eqnarray}
The corresponding free energy density equals

\begin{eqnarray}
F &=&\frac{1}{\beta }\stackunder{m}{\sum }\frac{1}{\left( 2\pi \right) ^{3}}%
\int d\mathbf{p}\frac{1}{2}Tr\left[ \widehat{S}(\mathbf{p},\varepsilon _{m})%
\widehat{\Delta }\widehat{S}^{T}(-\mathbf{p},-\varepsilon _{m})\widehat{%
\overline{\Delta }}.\right.  \nonumber \\
&&\left. .\left\{ \frac{1}{1+\widehat{S}(\mathbf{p},\varepsilon _{m})%
\widehat{\Delta }\widehat{S}^{T}(-\mathbf{p},-\varepsilon _{m})\widehat{%
\overline{\Delta }}}-\int\limits_{0}^{1}d\alpha \frac{1}{1+\alpha \widehat{S}%
(\mathbf{p},\varepsilon _{m})\widehat{\Delta }\widehat{S}^{T}(-\mathbf{p}%
,-\varepsilon _{m})\widehat{\overline{\Delta }}}\right\} \right] .\qquad
\label{12}
\end{eqnarray}

In order to study the quark-antiquark pairing we use the interaction
Lagrangian in the form (4) and start from the partition function

\begin{eqnarray}
Z &=&\int [D\psi ][D\overline{\psi }]\exp \left\{ -\int dx\overline{\psi }%
^{A}(x)D_{A}^{B}\psi _{B}(x)\right\}  \nonumber \\
&&.\exp \left\{ \frac{1}{2}\int dx\overline{\psi }^{A}(x)\psi
_{B}(x)U_{AC}^{BD}\hspace{0in}\overline{\psi }^{C}(x)\psi _{D}(x)\right\} .
\label{13}
\end{eqnarray}
Introducing the hermitian composite fields $\Phi _{B}^{A}(x)$ as well as the
functional integral

\begin{equation}
Z_{0}^{\Phi }=\int [D\Phi ]\exp \left\{ -\frac{1}{2}\int dx\Phi
_{B}^{A}(x)U_{AC}^{BD}\hspace{0in}\Phi _{D}^{C}(x)\right\}  \label{14}
\end{equation}
and applying the corresponding Hubbard-Stratonovich transformation, we
rewrite $Z$ in the form of the functional integral over these new composite
fields

\begin{equation}
Z=\frac{Z_{0}}{Z_{0}^{\Phi }}\int [D\Phi ]\exp \left\{ S_{\text{eff}}[\Phi
]\right\} .  \label{15}
\end{equation}
The field equations

\begin{equation}
\frac{\delta S_{\text{eff}}[\Phi ]}{\delta \Phi _{B}^{A}(x)}=0  \label{16}
\end{equation}
are the dynamical equations for the quark-antiquark pairing. They have the
explicit form

\begin{equation}
\Delta _{C}^{D}\left( x\right) =U_{CA}^{DB}G_{B}^{A}\left( x,x\right)
\label{17}
\end{equation}
where

\begin{equation}
\Delta _{C}^{D}\left( x\right) =U_{CA}^{DB}\Phi _{B}^{A}\left( x\right)
\label{18}
\end{equation}
and $G_{A}^{B}\left( x,y\right) $ are determined by the Schwinger-Dyson
equation

\begin{equation}
G_{A}^{B}\left( x,y\right) =S_{A}^{B}\left( x-y\right) -\int
dzS_{A}^{C}\left( x-z\right) \Delta _{C}^{D}\left( z\right) G_{D}^{B}\left(
z,y\right) .  \label{19}
\end{equation}
It is easy to verify that $G_{A}^{B}\left( x,y\right) $ are the two-point
Green functions of the quarks in the presence of the quark-quark pairing. In
the case of the constant order parameters

\[
\Delta _{A}^{B}=const 
\]
the Green functions $G_{A}^{B}\left( x,y\right) $ depend only on the
coordinate difference $x-y$. Their Fourier transforms $\widetilde{G}%
_{A}^{B}\left( \mathbf{p},\varepsilon _{m}\right) $ are determined by the
algebraic equation

\begin{equation}
\widetilde{G}_{A}^{B}\left( \mathbf{p},\varepsilon _{m}\right) =\widetilde{S}%
_{A}^{B}\left( \mathbf{p},\varepsilon _{m}\right) -\widetilde{S}%
_{A}^{C}\left( \mathbf{p},\varepsilon _{m}\right) \Delta _{C}^{D}\widetilde{G%
}_{D}^{B}\left( \mathbf{p},\varepsilon _{m}\right) .  \label{20}
\end{equation}
Introducing the matrices $\widehat{G}\left( \mathbf{p},\varepsilon
_{m}\right) \,$and $\widehat{\Delta }$ with the elements $\widetilde{G}%
_{A}^{B}\left( \mathbf{p},\varepsilon _{m}\right) $ and $\Delta _{C}^{D}$,
we obtain the matrix equation

\begin{equation}
\frac{1}{\widehat{G}\left( \mathbf{p},\varepsilon _{m}\right) }=\frac{1}{%
\widehat{S}\left( \mathbf{p},\varepsilon _{m}\right) }+\widehat{\Delta }.
\label{21}
\end{equation}
The free energy density of the system equals

\begin{eqnarray}
F\left[ \Delta \right] &=&\frac{1}{\beta }\stackunder{m}{\sum }\frac{1}{%
\left( 2\pi \right) ^{3}}\int d\mathbf{p}Tr\left[ \widehat{\Delta }\,%
\widehat{S}\left( \mathbf{p},\varepsilon _{m}\right) \right.  \nonumber \\
&&\left. .\left\{ \frac{1}{2}\frac{1}{1+\widehat{\Delta }\left( \mathbf{p}%
,\varepsilon _{m}\right) \widehat{S}\left( \mathbf{p},\varepsilon
_{m}\right) }-\int\limits_{0}^{1}d\alpha \frac{1}{1+\alpha \widehat{\Delta }%
\left( \mathbf{p},\varepsilon _{m}\right) \widehat{S}\left( \mathbf{p}%
,\varepsilon _{m}\right) }\right\} \right] .\qquad  \label{22}
\end{eqnarray}

%TCIMACRO{\TeXButton{TeX field}{\vskip 0.5cm}}
%BeginExpansion
\vskip 0.5cm%
%EndExpansion
For the study of the quark-quark pairing due to the gluon exchange we start
from the partition function in the form

\begin{eqnarray}
Z &=&\int [D\psi ][D\overline{\psi }]\exp \left\{ -\int dx\overline{\psi }%
^{A}(x)D_{A}^{B}\psi _{B}(x)\right\}  \nonumber \\
&&.\exp \left\{ \frac{1}{2}\int dx\int dy\overline{\psi }^{A}(x)\overline{%
\psi }^{C}(y)V_{CA}^{BD}\hspace{0in}\left( x-y\right) \psi _{D}(y)\psi
_{B}(x)\right\} ,  \label{23}
\end{eqnarray}
where

\begin{eqnarray}
V_{CA}^{BD}\hspace{0in}\left( x-y\right) &=&-\frac{g^{2}}{4\pi ^{2}}
\sum\limits_{I}\left( \gamma _{\mu }\otimes \lambda _{I}\right)
_{A}^{B}\left( \gamma _{\mu }\otimes \lambda _{I}\right) _{C}^{D}\frac{1}{%
\left( x-y\right) ^{2}},  \nonumber \\
\left( \gamma _{\mu }\otimes \lambda _{I}\right) _{A}^{B} &=&\left( \gamma
_{\mu }\right) _{\alpha }^{\beta }\left( \lambda _{I}\right) _{a}^{b}\delta
_{i}^{j},\quad \sum\limits_{I}\left( \lambda _{I}\right) _{a}^{b}\left(
\lambda _{I}\right) _{c}^{d}=\frac{1}{2}\left[ \delta _{c}^{b}\delta
_{a}^{d}-\frac{1}{N_{c}}\delta _{a}^{b}\delta _{c}^{d}\right] ,\quad
\label{24}
\end{eqnarray}
$\lambda _{I\text{ }}$ are the Gell-Mann matrices of the color symmetry
group. \noindent In order to describe the diquark systems we introduce some
bi-local composite fields $\Phi _{CD}\left( y,x\right) ,\overline{\Phi }%
^{AB}\left( x,y\right) .$ Consider the functional integral

\begin{equation}
Z_{0}^{\Phi ,\,\overline{\Phi }}=\int [D\Phi ][D\overline{\Phi }]\exp
\left\{ -\frac{1}{2}\int dx\int dy\overline{\Phi }^{AC}(x,y)V_{CA}^{BD}%
\hspace{0in}\left( x-y\right) \Phi _{DB}(y,x)\right\} .  \label{25}
\end{equation}
Applying the corresponding Hubbard-Stratonovich transformation, we rewrite
the partition function in the form

\begin{equation}
Z=\frac{Z_{0}}{Z_{0}^{\Phi ,\overline{\Phi }}}\int [D\Phi ][D\overline{\Phi }%
]\exp \left\{ S_{\text{eff}}[\Phi ,\overline{\Phi }]\right\} .  \label{26}
\end{equation}
The field equations

\begin{equation}
\frac{\delta S_{\text{eff}}[\Phi ,\overline{\Phi }]}{\delta \overline{\Phi }%
^{AC}(x,y)}=0  \label{27}
\end{equation}
are the dynamical equations for the quark-quark pairing. The bi-local
composite fields $\Phi _{DB}(y-x),\,\overline{\Phi }^{AC}(x-y)$ depending
only on the coordinate difference, or the linear combinations

\begin{equation}
\Delta _{CA}\left( y-x\right) =V_{CA}^{BD}\hspace{0in}\left( x-y\right) \Phi
_{DB}(y-x),\quad \overline{\Delta }^{BD}\left( x-y\right) =V_{CA}^{BD}%
\hspace{0in}\left( x-y\right) \overline{\Phi }^{AC}(x-y),\qquad \qquad
\label{28}
\end{equation}
are the order parameters of the superconducting phase transition. Denote $%
\widetilde{\Delta }_{CA}\left( \mathbf{p},\varepsilon _{m}\right) $ and $%
\widetilde{\overline{\Delta }}^{BD}\left( \mathbf{p},\varepsilon _{m}\right) 
$ their Fourier transforms and $\widehat{\Delta }\left( \mathbf{p}%
,\varepsilon _{m}\right) $,$\widehat{\overline{\Delta }}\left( \mathbf{p}%
,\varepsilon _{m}\right) $ the matrices with these elements. Then the field
equations (27) become the extended BCS equations for the color
superconductivity induced by the gluon exchange

\begin{eqnarray}
\widetilde{\Delta }_{CA}\left( \mathbf{p},\varepsilon _{m}\right)
&=&-g^{2}\sum\limits_{I}\left( \gamma _{\mu }\otimes \lambda _{I}\right)
_{A}^{B}\left( \gamma _{\mu }\otimes \lambda _{I}\right) _{C}^{D}.\frac{1}{%
\beta }\sum\limits_{n}\frac{1}{\left( 2\pi \right) ^{3}}\int d\mathbf{q} 
\nonumber \\
&&\frac{1}{\left( \varepsilon _{n}-\varepsilon _{m}\right) ^{2}+\left( 
\mathbf{p}-\mathbf{q}\right) ^{2}}\left[ \frac{1}{1+\widehat{S}\left( 
\mathbf{q},\varepsilon _{n}\right) \widehat{\Delta }\left( \mathbf{q}%
,\varepsilon _{n}\right) \widehat{S}^{T}\left( -\mathbf{q},-\varepsilon
_{n}\right) \widehat{\overline{\Delta }}\left( \mathbf{q},\varepsilon
_{n}\right) }\right.  \nonumber \\
&&\left. .\widehat{S}\left( \mathbf{q},\varepsilon _{n}\right) \widehat{%
\Delta }\left( \mathbf{q},\varepsilon _{n}\right) \widehat{S}^{T}\left( -%
\mathbf{q},-\varepsilon _{n}\right) _{{}}\right] _{DB}\qquad \qquad
\label{29}
\end{eqnarray}
The free energy density of the system equals

\begin{eqnarray}
F\left[ \Delta ,\overline{\Delta }\right] &=&\frac{1}{\beta }\sum\limits_{m}%
\frac{1}{\left( 2\pi \right) ^{3}}\int d\mathbf{p}Tr\left[ \widehat{S}\left( 
\mathbf{p},\varepsilon _{m}\right) ^{{}}\widehat{\Delta }\left( \mathbf{p}%
,\varepsilon _{m}\right) \widehat{S}^{T}\left( -\mathbf{p},-\varepsilon
_{m}\right) \widehat{\overline{\Delta }}\left( \mathbf{p},\varepsilon
_{m}\right) _{{}}\right.  \nonumber \\
&&\left\{ \left. \frac{1}{1+\widehat{S}\left( \mathbf{p},\varepsilon
_{m}\right) \widehat{\Delta }\left( \mathbf{p},\varepsilon _{m}\right) 
\widehat{S}^{T}\left( -\mathbf{p},-\varepsilon _{m}\right) \widehat{%
\overline{\Delta }}\left( \mathbf{p},\varepsilon _{m}\right) }-\right.
\right.  \label{30} \\
&&\left. \left. -\int\limits_{0}^{1}d\alpha \frac{1}{1+\alpha \widehat{S}%
\left( \mathbf{p},\varepsilon _{m}\right) \widehat{\Delta }\left( \mathbf{p}%
,\varepsilon _{m}\right) \widehat{S}^{T}\left( -\mathbf{p},-\varepsilon
_{m}\right) \widehat{\overline{\Delta }}\left( \mathbf{p},\varepsilon
_{m}\right) }\right\} \right] .  \nonumber
\end{eqnarray}

In order to study the quark-antiquark pairing we use the partition function
(23) in another form

\begin{eqnarray}
Z &=&\int [D\psi ][D\overline{\psi }]\exp \left\{ -\int dx\overline{\psi }%
^{A}(x)D_{A}^{B}\psi _{B}(x)\right\}  \nonumber \\
&&.\exp \left\{ \frac{1}{2}\int dx\int dy\overline{\psi }^{A}(x)\psi
_{B}(y)U_{AC}^{BD}\hspace{0in}\left( x-y\right) \overline{\psi }^{C}(y)\psi
_{D}(x)\right\} ,  \label{31}
\end{eqnarray}
\[
U_{AC}^{BD}\left( x-y\right) =-V_{CA}^{DB}\left( x-y\right) . 
\]
The quark-antiquark systems are described by the bi-local hermitian
composite fields $\Phi _{B}^{A}\left( x,y\right) $. Introducing the
functional integral

\begin{equation}
Z_{0}^{\Phi }=\int [D\Phi ]\exp \left\{ -\frac{1}{2}\int dx\int dy\Phi
_{B}^{A}(x,y)U_{AC}^{BD}\hspace{0in}\left( x-y\right) \Phi
_{D}^{C}(y,x)\right\}  \label{32}
\end{equation}
and applying the corresponding Hubbard-Stratonovich transformation, we
rewrite $Z$ in the form

\begin{equation}
Z=\frac{Z_{0}}{Z_{0}^{\Phi }}\int [D\Phi ]\exp \left\{ S_{\text{eff}}[\Phi
]\right\} .  \label{33}
\end{equation}
The field equations

\begin{equation}
\frac{\delta S_{\text{eff}}[\Phi ]}{\delta \Phi _{D}^{C}\left( y,x\right) }=0
\label{34}
\end{equation}
are the dynamical equations for the quark-antiquark pairing

\begin{equation}
\Delta _{C}^{D}\left( y,x\right) =U_{CA}^{DB}\hspace{0in}\left( x-y\right)
G_{B}^{A}\left( y,x\right) ,  \label{35}
\end{equation}
where $G_{B}^{A}\left( y,x\right) $ are determined by the Schwinger-Dyson
equations

\begin{equation}
G_{B}^{A}\left( y,x\right) =S_{A}^{B}\left( y-x\right) -\int dx^{\prime
}\int dy^{\prime }S_{B}^{A^{\prime }}\left( y-y^{\prime }\right) \Delta
_{A^{\prime }}^{B^{\prime }}\left( y^{\prime },x^{\prime }\right)
G_{B^{\prime }}^{A}\left( x^{\prime }-x\right)  \label{36}
\end{equation}
and

\begin{equation}
\Delta _{A}^{B}\left( x,y\right) =U_{AC}^{BD}\left( x-y\right) \Phi
_{D}^{C}\left( y,x\right) .  \label{37}
\end{equation}
$G_{B}^{A}\left( y,x\right) $ are the two point Green functions of quarks in
the presence of the quark-antiquark pairing.\noindent In the special class
of the functions $\Delta _{A}^{B}\left( x-y\right) $ depending only on the
coordinate difference these functions are the order parameters of the
system. Denote $\widetilde{\Delta }_{A}^{B}\left( \mathbf{p},\varepsilon
_{m}\right) $ their Fourier transforms and $\widehat{\Delta }\left( \mathbf{p%
},\varepsilon _{m}\right) $ the matrix with these elements. The dynamical
equations (35) are 
\begin{eqnarray}
\widetilde{\Delta }_{C}^{D}\left( \mathbf{p},\varepsilon _{m}\right)
&=&g^{2}\sum\limits_{I}\left( \gamma _{\mu }\otimes \lambda _{I}\right)
_{A}^{D}\left( \gamma _{\mu }\otimes \lambda _{I}\right) _{C}^{B}  \nonumber
\\
&&\frac{1}{\beta }\sum\limits_{n}\frac{1}{\left( 2\pi \right) ^{3}}\int d%
\mathbf{q}\frac{1}{\left( \varepsilon _{n}-\varepsilon _{m}\right)
^{2}+\left( \mathbf{p}-\mathbf{q}\right) ^{2}}\widetilde{G}_{B}^{A}\left( 
\mathbf{q},\varepsilon _{n}\right) ,  \label{38}
\end{eqnarray}
and the Schwinger-Dyson equations become

\begin{equation}
\frac{1}{\widehat{G}\left( \mathbf{q},\varepsilon _{m}\right) }=\frac{1}{%
\widehat{S}\left( \mathbf{q},\varepsilon _{m}\right) }+\widehat{\Delta }%
\left( \mathbf{q},\varepsilon _{m}\right) .  \label{39}
\end{equation}
The free energy density of the system equals 
\begin{eqnarray}
F\left[ \Delta \right] &=&\frac{1}{\beta }\stackunder{m}{\sum }\frac{1}{%
\left( 2\pi \right) ^{3}}\int d\mathbf{p}Tr\left[ \widehat{\Delta }\left( 
\mathbf{p},\varepsilon _{m}\right) \widehat{S}\left( \mathbf{p},\varepsilon
_{m}\right) \right.  \nonumber \\
&&\left. \left\{ \frac{1}{2}\frac{1}{1+\widehat{\Delta }\left( \mathbf{p}%
,\varepsilon _{m}\right) \widehat{S}\left( \mathbf{p},\varepsilon
_{m}\right) }-\int\limits_{0}^{1}d\alpha \frac{1}{1+\alpha \widehat{\Delta }%
\left( \mathbf{p},\varepsilon _{m}\right) \widehat{S}\left( \mathbf{p}%
,\varepsilon _{m}\right) }\right\} \right] .\qquad  \label{40}
\end{eqnarray}

The existence of the non-vanishing order parameters $\Delta _{CA}$, $%
\overline{\Delta }^{BD}$ or $\Delta _{CA}\left( y-x\right) $, $\overline{%
\Delta }^{BD}\left( x-y\right) $ which cannot be the singlets of the color
symmetry group would mean the spontaneous breaking of the color symmetry$%
^{\left[ 29\right] }$.

The physical vacuum in QCD is the limiting case of the system with vanishing
density ($\mu =0$) at vanishing temperature ($T=0$). In this limit the order
parameters $\Delta _{DC}$, $\overline{\Delta }^{AB}$ and $\Delta _{B}^{A}$
are the vacuum expectation values of the corresponding quantum composite
fields - the new Higgs fields.

The functional integral method was applied also to the study of the binding
of three quarks due to some effective six-fermion (non-local, in general)
interactions of quarks. The systems of dynamical equations for the composite
fermionic quantum triquark fields $\Psi _{ABC}$ and $\overline{\Psi }^{ABC}$
were derived. The constant solutions of these equations in the case of the
direct six-fermion couplings of quarks or their solutions depending only on
the coordinate difference in the case of the non-local effective six-fermion
interactions are the anticommuting order parameters.\newpage 

%TCIMACRO{\TeXButton{TeX field}{\vskip 0.4cm}}
%BeginExpansion
\vskip 0.4cm%
%EndExpansion
%TCIMACRO{\TeXButton{References.}{{\Large \bf Acknowledgements.}}}
%BeginExpansion
{\Large \bf Acknowledgements.}%
%EndExpansion

%TCIMACRO{\TeXButton{TeX field}{\vskip 0.3cm}}
%BeginExpansion
\vskip 0.3cm%
%EndExpansion
The author would like to express his sincere appreciation to the National
Natural Sciences Council of Vietnam for the support to this work.%
%TCIMACRO{\TeXButton{TeX field}{\vskip 0.4cm}}
%BeginExpansion
\vskip 0.4cm%
%EndExpansion

%TCIMACRO{\TeXButton{References.}{{\Large \bf References.}}}
%BeginExpansion
{\Large \bf References.}%
%EndExpansion

\begin{itemize}
\item[{\lbrack 1].}]  B. C. Barrois, \textit{Nucl. Phys.} \textbf{B129}
(1977) 390.

\item[{\lbrack 2].}]  C. Frautshi, \textit{Asymptotic Freedom and Color
Superconductivity in Dense Quark Matter, in: Proceedings of the Workshop on
Hadronic Matter at Extreme Energy Density}, Ed., N. Cabibbo, Erice, Italy
(1978).

\item[{\lbrack 3].}]  D. Bailin and A. Love, \textit{Nucl. Phys.} \textbf{%
B190} (1981) 175; \textbf{B190} (1981) 751; \textbf{B205} (1982) 119.

\item[{\lbrack 4].}]  J. F. Donoghue and K. S. Sateesh, \textit{Phys. Rev.} 
\textbf{D38} (1988) 360.

\item[{\lbrack 5].}]  M. Iwasaki and T. Iwado, \textit{Phys. Lett.} \textbf{%
B350} (1995) 163.

\item[{\lbrack 6].}]  M. Alford, K. Rajagopal and F. Wilczek , \textit{Phys.
Lett. }\textbf{B422} (1998) 247; \textit{Nucl. Phys},. \textbf{B537} (1999)
443.

\item[{\lbrack 7].}]  T. Sch\"{a}fer and F. Wilczek, \textit{Phys. Rev. Lett.%
} \textbf{82} (1999) 3956; \textit{Phys. Lett.} \textbf{B450} (1999) 325.

\item[{\lbrack 8].}]  R. Rapp, T. Sch\"{a}fer, E. Shuryak and M. Velkovsky, 
\textit{Phys. Rev. Lett.},\textbf{\ 81} (1998) 53.

\item[{\lbrack 9].}]  N.Evans, S. Hsu and M. Schwetz, \textit{Nucl. Phys.} 
\textbf{B551} (1999) 275; \textit{Phys. Lett.} \textbf{B449} (1999) 281.

\item[{\lbrack 10].}]  D. T. Son, \textit{Phys. Rev.} \textbf{D59 }(1998)
094019.

\item[{\lbrack 11].}]  G. W. Carter and D. Diakonov, \textit{Phys. Rev}. 
\textbf{D60 }(1999) 01004.

\item[{\lbrack 12].}]  R. Rapp, T. Sch\"{a}fer, E. Shuryak and M. Velkovsky,
preprint IASSNS-HEP-99/40, Princeton, 1999, hep-ph/9904353.

\item[{\lbrack 13].}]  A. Chodos, H. Minakata and F. Cooper, \textit{Phys.
Lett}. \textbf{B449} (1999) 260.

\item[{\lbrack 14].}]  R. D. Pisarski and D. H. Rischke, \textit{Phys. Rev}. 
\textbf{D60} (1999) 094013.

\item[{\lbrack 15].}]  T Sch\"{a}fer and F. Wilczek, \textit{Phys. Rev}. 
\textbf{D60} (1999) 074014, 114033.

\item[{\lbrack 16].}]  V. A. Miransky, I. A. Shovkovy and L. C.
Wijewardhana, \textit{\ Phys. Lett. }\textbf{B468} (1999) 270.

\item[{\lbrack 17].}]  R. Casalbuoni and R. Gatto, \textit{Phys. Lett}. 
\textbf{B464} (1999) 111.

\item[{\lbrack 18].}]  D. K. Hong, M. Rho and I. Zahed, \textit{Phys. Lett}. 
\textbf{B468} (1999) 261.

\item[{\lbrack 19].}]  M. Rho, E. Shuryak, A. Wirzba and I. Zahed, \textit{%
hep-/0001104}.

\item[{\lbrack 20].}]  C. Manuel and M. H. G. Tytgat, \textit{hep-ph/0001095.%
}

\item[{\lbrack 21].}]  W. Brown, J. T. Liu and H. C. Ren, \textit{%
hep-ph/0003199.}

\item[{\lbrack 22].}]  J. Berges and K. Rajagopal, \textit{Nucl. Phys.} 
\textbf{B358} (1999) 215.

\item[{\lbrack 23].}]  M. Harada and A. Shibata, \textit{Phys. Rev.} \textbf{%
D59} (1998) 014010.

\item[{\lbrack 24].}]  R. D. Pisarski and D. H. Rischke, \textit{Phys. Rev.
Lett}. \textbf{83} (1999) 37.

\item[{\lbrack 25].}]  Nguyen Van Hieu, \textit{Basics of Fuctional Integral
Technique in Quantum Field Theory of Many-Body Systems, VNUH Pub}., Hanoi,
1999.

\item[{\lbrack 26].}]  Nguyen Van Hieu and Le Trong Tuong, \textit{Commun.
Phys.} \textbf{8} (1998) 129.

\item[{\lbrack 27].}]  Nguyen Van Hieu, Nguyen Hung Son, Ngo Van Thanh and
Hoang Ba Thang, \textit{Adv . Nat. Sci}. \textbf{1} (2000) 61, \textit{%
hep-ph/0001251.}

\item[{\lbrack 28].}]  Nguyen van Hieu, Hoang Ngoc Long, Nguyen Hung Son and
Nguyen Ai Viet, \textit{Proc. 24th Conference in Theoretical Physics, Samson
19-21 August 1999, p.1}, \textit{hep-ph/0001234.}

\item[{\lbrack 29].}]  C. Wetterich, \textit{hep-ph/0008150.}
\end{itemize}

\end{document}